\documentstyle[preprint,aps,epsfig]{revtex}

\begin{document}

\title{Mode Repulsion and Mode Coupling in Random Lasers}

\author{H. Cao$^a$, Xunya Jiang$^b$, Y. Ling$^a$, J.Y. Xu$^a$, C.M.
 Soukoulis$^b$}

\address{$^a$ Department of Physics and Astronomy, Northwestern University,
 \\ 2145 Sheridan Road, Evanston, IL 60208-3112}

\address{$^b$ Ames Laboratory-USDOE and Department of Physics and
 Astronomy,\\ Iowa State University, Ames, IA 50011}


\maketitle

\begin{abstract}
We studied experimentally and theoretically the interaction of lasing modes in random media. In a homogeneously broadened gain medium, cross gain saturation leads to spatial repulsion of lasing modes. In an inhomogeneously broadened gain medium, mode repulsion occurs in the spectral domain. Some lasing modes are coupled through photon hopping or electron absorption and reemission. Under pulsed pumping, weak coupling of two modes leads to synchronization of their lasing action. Strong coupling of two lasing modes results in anti-phased oscillations of their intensities.

\end{abstract}



\newpage

A random laser is a special kind of laser whose feedback is due to disorder-induced scattering, as opposed to reflection of a conventional laser. Over the past few years, there have been many studies on random lasers with coherent feedback. \cite{cao,fro,pat,cao1,jia,cao2,pol,seb,bur,jia1,sou,bur1,pol1,apa} One important aspect of a random laser is the interaction of the lasing modes, which has not been well understood. \cite{jia,bri,mis} As previously pointed out \cite{jia1}, adding optical gain to a random medium creates a new path to study wave transport and localization.  Because the modes of coherent random lasers are eigenstates of the random systems, the interaction of the lasing modes reflects the interaction among the eigenstates. So far most studies on eigenstates of random systems are focused on statistical properties such as the density of states and the distribution of spectral spacing between neighboring eigenstates. \cite{bee} However, there has been no detailed study on the interaction between two eigenstates in a disordered system, e.g., the spatial overlap of two eigenstates and the photon hoping between them. Such $microscopic$ information is crucial to understanding the behavior of light in the random system. By studying the interaction of random lasing modes, we will gain some knowledge of the interaction among the eigenstates of a disordered system. Furthermore, this study will provide an understanding of the escape channels for photons in an eigenstate, that is important to the localization theory.

In this letter, we present experimental and theoretical studies on the interaction of modes in random lasers with coherent feedback. We find the interaction is rather complicated. On one hand, most lasing modes repel each other either spatially or spectrally. On the other hand, some lasing modes are coupled. Energy exchange between the coupled modes leads to cooperated lasing phenomenon. 

The spatial repulsion of lasing modes was first observed in the numerical
 simulation of random lasers. \cite{jia}  It results from gain competition
 and spatial localization of the modes in an active random medium. The gain
 spectrum is assumed to be homogeneously broadened. In a finite-sized random
 medium, the mode repulsion leads to a saturation in the number of lasing
 modes. \cite{jia} Experimentally we observed the number of lasing modes in
 micropatterned ZnO polycrystalline films saturates at high pumping
 intensity. \cite{lin} The saturated number of lasing modes increases linearly with the sample volume.

In random media with inhomogeneously broadened gain spectra, we observed a
 different kind of mode repulsion. The lasing modes repel each other in the
 spectral domain instead of the spatial domain. The disordered samples used
 in our experiment are poly(methyl methacrylate) (PMMA) sheets containing
 rhodamine 640 perchlorate dye and titanium dioxide (TiO$_2$)
 microparticles. The concentration of rhodamine 640 is about 10 mM. Each individual dye molecule's transition frequency is shifted by differing local strains and distortions in the immediate environment of the glassy host. This gives rise to a large inhomogeneous
 broadening of the dye transition, which is confirmed by good fitting of the absorption band tail with a Gaussian profile (the error of Gaussian curve fitting is ten times smaller than that of Lorentzian curve fitting). The TiO$_2$ particles have an average diameter of
 0.4 $\mu$m. The density of the microparticles is $\sim$ 1.4 $\times$
 10$^{12}$ cm$^{-3}$. The transport mean free path, measured in the coherent
 backscattering experiment, is 2.1 $\mu$m.  \cite{lin}

The rhodamine 640 dye molecules are optically excited by the second
 harmonics of a pulsed Nd:YAG laser (10 Hz repetition rate, 25 ps pulse
 width). The pump spot is about 50 $\mu$m in diameter at the sample surface.
 The spectrum of emission from the sample is measured by a 0.5-meter
 spectrometer with a liquid nitrogen cooled CCD array detector. The spectral
 resolution is 0.06 nm.
\begin{figure}
\centering
\vspace{-0.5in}
\epsfxsize=4in \epsfbox{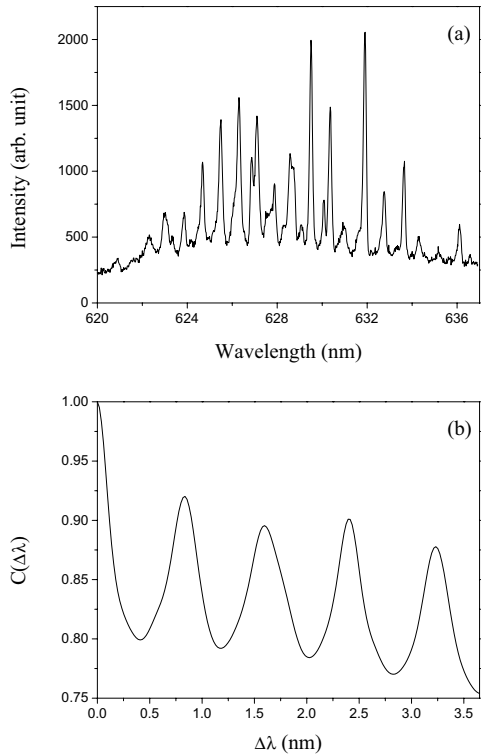}
\vspace{-0.2in}
\caption{(a) Spectrum of laser emission from the PMMA containing
 dye and microparticles. The incident pump pulse energy is 1.6 $\mu$J. (b)
 Normalized spectral correlation function $C(\Delta \lambda)$.}
\end{figure}
Below the lasing threshold, the emission spectrum is very broad. When the 
 pumping intensity exceeds the threshold, discrete narrow peaks emerge in 
 the emission spectrum. The number of lasing modes increases with the 
 pumping intensity. Eventually the spectral density of the lasing peaks 
 becomes saturated. Namely, the wavelength spacing between adjacent 
lasing 
 modes no longer decreases as the pumping intensity increases. New lasing
 peaks appear not in between the existing peaks but in the spectral region
 that is away from the gain maximum and has few lasing peaks. The lasing
 spectrum, shown in Fig. 1 (a), reveals that the lasing peaks are more or
 less regularly spaced in wavelength. We calculated the normalized spectral
 correlation function $C(\Delta \lambda) \equiv \int I(\lambda) \, I(\lambda
 + \Delta \lambda) d \lambda / \int [I(\lambda)]^2 d \lambda$, where
 $I(\lambda)$ is the emission intensity at wavelength $\lambda$.  Fig. 1 (b)
 shows the normalized spectral correlation function $C(\Delta \lambda)$
 obtained from the lasing spectrum in Fig. 1 (a). The periodic
 oscillation of $C(\Delta \lambda)$ with $\Delta \lambda$ confirms the
 lasing modes are evenly distributed in wavelength. The oscillation period gives the average separation between adjacent lasing modes, which is 0.8 nm. Note that the minimum of $C(\Delta \lambda)$ does not reach zero because the lasing peaks are on top of a 
 broad band of amplified spontaneous emission.  

Recently we developed a spectrally resolved speckle technique which gives
 the spatial profile of individual lasing modes at the sample surface.
 \cite{cao3} Using this technique, we find the lasing modes in Fig. 1 (a)
 have different sizes. The radius of the lasing modes varies from 100 $\mu$m
 to 150 $\mu$m, which is larger than the radius of the pump spot. To
 acquire gain, the lasing modes must overlap with the excitation area. This
 suggests most lasing modes spatially overlap with each other. Owing to a
 large inhomogeneous broadening of the dye molecules, lasing can occur
 simultaneously in modes that are spatially overlapped, as long as the
 frequency separation between the modes is larger than the homogeneous
 linewidth of the dye molecules. This is because a lasing mode 
 interacts only with the dye molecules whose transition frequency is in
 resonance with the lasing frequency, and depletes the population inversion
 of these dye molecules. As a result, a hole is burned in the gain spectrum,
 and its width is about twice of the homogeneous linewidth of the dye
 molecules. \cite{siegman} Nevertheless, the dye molecules, whose frequency
 is more than one homogeneous linewidth away from the frequency of the
 lasing mode, are essentially unaffected by this lasing mode. They provide
 optical gain for other modes in the same area. Thus the frequency spacing of the lasing modes is limited by the homogeneous linewidth of the dye molecules. At room
 temperature, the homogeneous linewidth of rhodamine 640 dye molecules
 in PMMA is $\sim$ 0.5 nm. \cite{elsc,bill,baer} The linewidth of the hole burned by a lasing mode is about 1 nm, close to the wavelength separation between adjacent lasing modes. This confirms the regular spectral spacing of the lasing modes is caused by the hole burning of the inhomogeneously broadened gain spectrum. The spectral repulsion of lasing modes can only be observed when the excitation area is smaller than the spatial extent of lasing modes. When the pump area is much larger than the size of lasing modes, there is no minimum frequency separation between adjacent lasing modes. This is because the lasing modes are not necessarily overlapped in space. If they are spatially separated, the lasing modes can have the same frequency. 

In addition to mode repulsion, we also observed mode coupling in random
 lasers. Despite most lasing modes in Fig. 1 (a) are well separated
 spectrally, we do see two or three lasing modes partially overlapped in the
 spectrum, e.g. the two lasing modes centered at 626.9 nm and 627.1 nm.
 Since their frequency separation is less than the homogeneous linewidth of
 the dye molecules, these two lasing modes interact with the same group of
 excited dye molecules. The gain competition would not allow them to lase 
 simultaneously unless they are coupled, i.e., there is an energy exchange
 between the two modes.

\begin{figure}
\centering
\vspace{-0.5in}
\epsfxsize=4in \epsfbox{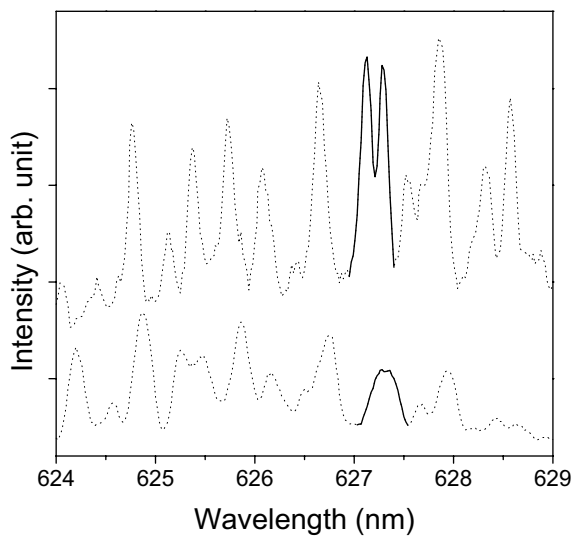}
\vspace{-0.2in}
\caption{Measured spectra of laser emission when the incident pump pulse energies are 0.61 $\mu$J and 0.98 $\mu$J. The spectra are shifted vertically for clarity. The solid lines highlight the growth of two coupled lasing modes. }
\end{figure}
Figure 2 shows the growth of two coupled lasing modes with the pumping
 intensity. Just above the lasing
 threshold, a relatively broad lasing peak appears in the emission spectrum.
 As the pumping intensity increases, this lasing peak is slightly
 blue-shifted and split into two narrow peaks. It seems initially the
 linewidth of the two lasing modes is larger than their spectral separation,
 thus they appear to be a single broad peak. As the optical gain increases,
 the lasing peaks become narrower. When their linewidth is reduced to below
 their spectral separation, these two lasing modes can be spectrally
 resolved, leading to a splitting of the initial peak. We measured the
 temporal evolution of emission from these two modes with a streak camera. Lasing in these two modes is synchronized, i.e., lasing starts and stops at the same time.

\begin{figure}
\centering
\vspace{-0.5in}
\epsfxsize=4in \epsfbox{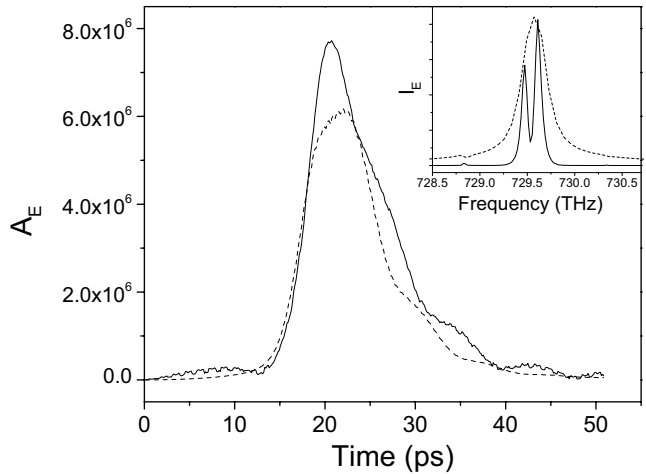}
\vspace{-0.2in}
\caption{Calculated amplitudes of electric field $A_E$ for two coupled lasing modes versus time. The number of cells is 100. The pumping rate $P_r = 4 \times 10^6$ s$^{-1}$. The pumping time is 15 ps. The inset are the calculated spectra for pumping rates of $2 \times 10^6$ s$^{-1}$ (dashed line) and $4 \times 10^6$ s$^{-1}$ (solid line). }
\end{figure}

To understand the coupling of the random laser modes, we performed numerical simulations using the time-dependent model for random lasers. \cite{jia,sou} The Maxwell's equations are coupled to the rate equations of electronic population. Because the coupled modes are within one homogeneous line and interact with the same group of molecules, we can assume the electronic energy levels are homogeneously broadened. The homogeneous linewidth is 5 THz. Temporal evolution of the electromagnetic field is obtained with the finite difference time domain (FDTD) method. Fourier transform is used to obtain the lasing spectra. The one-dimensional random system is made of binary layers with dielectric constants $\epsilon_1$ = 2.5 and $\epsilon_2$ = 5.29. The layer thickness $a=a_0 (1+W_a \gamma)$, $b=b_0 (1+W_b \gamma)$, where $a_0$ = 48 nm, $W_a$ = 0.1, $b_0$ = 80 nm, $W_b$ = 0.6, $\gamma$ is a random value in the range [-1,1]. In the inset of Fig. 3, we plot the calculated lasing spectra for a system of 100 pairs of binary layers at two pumping rates. The pumping starts at $t$ = 0 and stops at $t$ = 15 ps. At a pumping rate $P_r = 2 \times 10^6$ s$^{-1}$, the spectrum has one relatively broad peak. This peak is split into two narrow peaks at a pumping rate $P_r = 4 \times 10^6$ s$^{-1}$. Figure 3 shows the temporal evolution of the amplitude of the electric field $A_E$ for the two lasing peaks.  Laser pulses in these two modes rise at the same time. The pulse duration is nearly identical. These results agree with the experimental data.

The synchronization of the two lasing modes results from coupling. The quality factors of these two modes are different. Our previous study of random lasers demonstrates that lasing modes with different quality factors exhibit unsynchronized temporal behaviors. \cite{sou} This conclusion holds only for non-interacting modes. However, lasing modes in a random medium can be coupled through photons and electrons. Namely, photons can hop from one mode to another as long as these two modes have spatial and spectral overlap. In addition, photons in one mode can be absorbed by electrons in the medium, which subsequently remit photons into another mode. To be more specific, let us consider two coupled modes A and B. The quality factor of mode A is higher than that of mode B. After pumping starts, as optical gain increases,  the lasing threshold for mode A is reached first. The electromagnetic field of mode A increases quickly. When there are more photons in mode A than in mode B, there is a net energy flow from mode A to mode B through photon hopping or electron absorption and reemission. This seeding increases the photon number of mode B and accelerates lasing process in mode B. Thus mode A provides additional pumping to mode B. If the quality factor of mode B is not much lower than that of mode A, mode B can reach the lasing threshold while mode A still builds up. That is why we observe the two modes lase almost simultaneously. Similarly, when the pumping is stopped, photon number of mode B decreases faster than that of mode A due to higher loss rate of mode B. The difference of photon numbers in these two modes leads to a net energy flow from mode A to mode B. Thus mode B provides an additional leak channel for mode A, and photons in mode A are drained faster. As a result, the two modes stop lasing at almost the same time.     

Next we increased the quality factors of the coupled modes by enlarging the system to 120 pairs of binary layers. We also increased the pumping time to 25 ps to make the laser pulses longer. As shown in Fig. 4, the dynamics of the coupled modes is very different from that shown in Fig. 3. Laser light intensities in these two modes oscillate in time. Their oscillations are 180$^{\circ}$ out of phase. Namely, when one mode reaches the maximum intensity, the other mode is at the minimum intensity. These antiphased oscillations reveal energy exchange between the two coupled modes. From the numerical simulation, we find such oscillations occur only when the coupled modes have high quality factors and the pumping time is not too short. This is because photons must stay in the medium long enough so that they are transferred back and forth between the coupled modes before escaping the system. The oscillation period is inversely proportional to the energy exchange rate or the coupling constant. The coupling of the two modes depends on their spectral and spatial overlap. There are two coupling regimes. In the weak coupling regime, the coupling constant of the two modes is smaller than their decay rates. In the strong coupling regime, the coupling constant is larger than the decay rates of the two modes. The oscillations occur only in the strong coupling regime. In our current experiment with the three-dimensional (3D) random medium, the quality factors of the lasing modes are too low to reach the strong coupling regime. The antiphased oscillations of the coupled modes could not be observed. 
\begin{figure}
\centering
\vspace{-0.5in}
\epsfxsize=4in \epsfbox{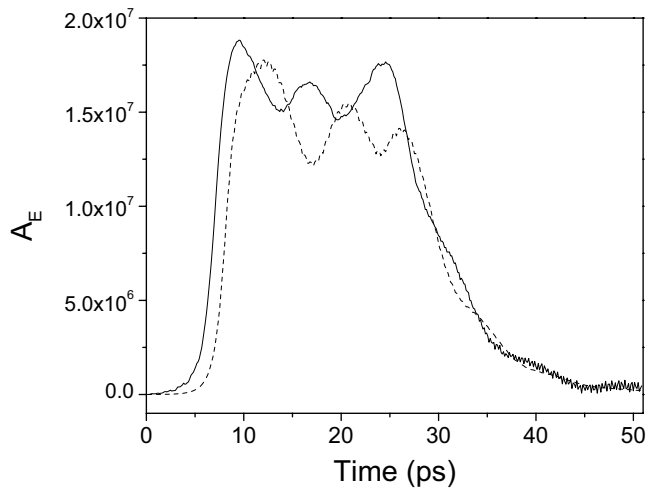}
\vspace{-0.2in}
\caption{Calculated amplitudes of electric field $A_E$ for two coupled lasing modes versus time. The number of cells is 120. The pumping time is 25 ps. The pumping rate $P_r = 4 \times 10^6$ s$^{-1}$.}
\end{figure}
To confirm the spectral repulsion of lasing modes, we simulated random laser with inhomogeneously broadened gain spectrum. The inhomogeneous linewidth is 150 THz, 30 times of the homogeneous linewidth. To increase the spatial overlap of the eigenstates, we reduce the randomness to $W_a$ = 0 and $W_b$ = 0.2 so that the modes are more extended spatially. The size of the system is increased to 700 pairs of binary layers ($\sim$ 100 $\mu$m). However, only the central part of the system with the length 30 $\mu$m is pumped. Since the lasing modes must overlap with the pumped region, they overlap with each other. At very high pumping rate, the lasing modes are almost regularly spaced in frequency. The spectral correlation function $C(\Delta \nu)$ exhibits periodic oscillation with $\Delta \nu$. The frequency separation of neighboring lasing modes is about 7 THz. It is much larger than the average frequency spacing of the eigenstates in the random system. This means most eigenstates do not lase even at very high pumping rate due to cross gain saturation. More specifically, there are about three eigenstates within one homogeneous line. Cross gain saturation allows only one of them to lase. Thus the frequency spacing of the lasing modes is determined by the homogeneous linewidth.

Finally we increased the pumping time in our simulation. All the phenomena presented above  persist. Despite our experiments were performed with short pump pulses, our numerical simulation results confirm that mode repulsion and mode coupling exist at longer pumping time.

In conclusion, we studied the interaction of lasing modes in random media. In a homogeneously broadened gain medium, cross gain saturation leads to spatial repulsion of lasing modes. In an inhomogeneously broadened gain medium, mode repulsion occurs in the spectral domain. Some lasing modes are coupled through photon hopping or electron absorption and remission. Under pulsed pumping, weak coupling of two modes leads to synchronization of their lasing action. Strong coupling of two lasing modes results in anti-phased oscillations of their intensities. 

Ames Laboratory is operated for the U.S. Department of Energy by Iowa State University under Contract No. W07405-Eng-82. This work is supported partially by National Science Foundation under Grant No. DMR-0093949. H.C. acknowledge support from the David and Lucille Packard Foundation and the Alfred P. Sloan Foundation.


\end{document}